\documentclass[lettersize,journal]{IEEEtran}
\usepackage{amsmath,amsfonts}
\usepackage{algorithmic}
\usepackage{algorithm}
\usepackage{array}
\usepackage{textcomp}
\usepackage{stfloats}
\usepackage{url}
\usepackage{verbatim}
\usepackage{graphicx}
\usepackage{cite}
\usepackage{booktabs}
\usepackage{amsmath,amsfonts}
\usepackage{algorithmic}
\usepackage{array}
\usepackage{textcomp}
\usepackage{stfloats}
\usepackage{subcaption}
\usepackage{url}
\usepackage{verbatim}
\usepackage{graphicx}
\usepackage{float}
\usepackage{titlesec}
\usepackage{amsthm}
\usepackage{mathtools}
\usepackage{gensymb}
\usepackage{hyperref}
\usepackage{amssymb}
\usepackage{amsmath}
\usepackage{dsfont}

\newcommand{\R}{\mathbb{R}}

\newtheorem{theorem}{Theorem}

\newtheorem{prop}[theorem]{Proposition}

\def\BibTeX{{\rm B\kern-.05em{\sc i\kern-.025em b}\kern-.08em
    T\kern-.1667em\lower.7ex\hbox{E}\kern-.125emX}}
\begin{document}

\title{Stochastic Analysis of Successive Interference Cancellation in a Narrow-Beam LEO Uplink}
\author{
   Ilari Angervuori,   and Risto Wichman \thanks{Ilari Angervuori and Risto Wichman are with the School of Electrical Engineering, Aalto University, Espoo, 02150, Finland. (email: ilari.angervuori@aalto.fi; risto.wichman@aalto.fi).}
 }


\IEEEpubid{}

\maketitle
\begin{abstract}
This paper investigates uplink coverage probability, i.e., the SIR distribution, and SIR order statistics at a low Earth orbit satellite base station (LEO BS) with narrow Gaussian antenna beams. The user equipments (UEs) connected to the satellite radio access network are distributed according to a Poisson point process (PPP) on the Earth's surface. We propose a novel framework for stochastic geometry analysis of the SIR using a Gaussian mixture shadowing model, derive tractable expressions for the mean, variance, and probability density function, and analyze the SIR distributions for the three strongest UEs under successive interference cancellation (SIC). Furthermore,  we propose a gamma distribution model for the interference distribution. We show that SIC enables each LEO BS to serve multiple UEs per beam cell, improving user fairness while preserving the network throughput. The observations imply design recommendations for next-generation non-orthogonal multiple access LEO networks. The analytical insights are validated against Monte Carlo simulations.
\end{abstract}


\section{Introduction and motivation}

Successive interference cancellation (SIC) has the potential to improve the performance of non-orthogonal multiple access (NOMA) communication systems and is a hot issue for future wireless communications~\cite{Flizikowski2023NOMASurvey}.

Stochastic geometry analysis of the signal-to-interference ratio (SIR) under interference cancellation (and signal combination) has been extensively studied for terrestrial networks, with one of the pioneering works being \cite{7305791}. In non-terrestrial networks (NTN), a SIC scheme was studied in~\cite{11030605} using stochastic geometry in a NOMA satellite network. However, while NOMA stochastic geometry frameworks have been developed for NTN~\cite{ShirzadianGilan2025HSUAV, Gao2025SatMaritime}, a comprehensive framework for the analysis of SIR order statistics and SIC is, to the best of our knowledge, yet to be explored in Earth orbit (LEO) networks.

Notably, for the LEO network, the order statistics of the SIR process are characterized by a Poisson-Dirichlet distribution PD$(0, \cdot)$, in contrast to terrestrial networks with a singular path loss, where the corresponding process is PD$(\cdot,0)$ \cite{7305791}. That is, the order statistics are characterized by a gamma process, making the analysis fundamentally distinct from that of terrestrial networks, where the gamma process does not apply. Furthermore, in contrast to terrestrial networks, the spatial density of the transmitters has a substantial effect on the SIR in the satellite network. On the contrary, the spatial path loss exponent has a negligible effect in the case of narrow-beam LEO BSs.

In this paper, we study the SIR under the SIC of user equipments (UEs) at a typical LEO satellite base station (BS) beam in an urban shadowing environment. We utilize the Gaussian Mixture shadowing model, as proposed in \cite{modelinguplink} and \cite{9684552}, and approximate it with a \textit{mixture exponential} distribution to simplify the analysis. Furthermore, we propose a planar system model. 
The analytical results are verified by comparing them with more realistic Monte Carlo-simulated performance metrics in a realistic system model with a spherical Earth and Gaussian mixture shadowing.
We demonstrate that, while maintaining good performance, link quality can be made more stable by increasing the number of UEs served by each LEO BS beam, while mitigating interference with the SIC.

\section{System model}

\begin{table}
   \captionsetup{size=footnotesize}
   \caption{Glossary of principal symbols and the values used in the simulated numerical results. We denote proportionality ``$\propto$'' or equality ``$=$'' to a variable. For a dimensional number, we denote the units ``SI'' and ``[non-SI]''}
     \label{table:parameters}
     \begin{center}
    \begin{tabular}{c|p{4.5cm}|p{1.9cm}}
      \toprule
      Symbol& Explanation &Values and units
      \\ 
      \hline 
      $h$ & Receiving LEO BS altitudes. &$1000$ km  \\
      $\alpha$ &Power path loss exponent (negligible effect on the SIR).& $2$\\
      $\epsilon$& The typical LEO BS beam center elevation angle.&$({\pi}/{6},{\pi}/{2})\text{ rad}\break=(30,90)[\degree]$\\
      $\lambda$ & The average number of UE transmitters inside a unit area. &$1.44 \times \hfill \break 10^{-3}\text{/km}^2$\\
      $p_{\text{LoS}}$& LoS probability; $p_{\text{LoS}} \propto \sin(\epsilon)$. & $(0.992,0.493)$\\
      $\mu_{\text{LoS}}$& Mean of the LoS component of the Gaussian mixture shadow fading model. & $0$ [dB] \\
      $\mu_{\text{NLoS}}$& Mean of the NLoS component. & $-26$ [dB] \\
      $\sigma_{\text{LoS}}$& Standard deviation of the LoS component. & $4$ [dB] \\
      $\sigma_{\text{NLoS}}$& Standard deviation of the NLoS component. & $6$ [dB] \\
      $\varphi_{\text{RX}}$ & Half-width of the $-3$ dB beams: $\varphi_{\text{RX}}=\varphi_{\text{RX}}(\epsilon) \propto \sin^{3/2}(\epsilon)$.&$(0.010,0.028)= \hfill(0.6,1.6) [\degree] \hfill $ \\      
      $\kappa$& Average number of UEs inside a $-3$ dB beam footprint (a cell); $\kappa =\hfill\pi \lambda \left({\varphi_{\text{RX}}}h/{\sin^2(\epsilon)}\right)^2$.& $\{3.53,7.04\}$ \\
      $\upsilon$&Fraction of the transmitters that are not entirely attenuated, i.e., the effective UEs; $\upsilon \propto \sin(\epsilon)$. &$\{0.85,0.426\}$ \\      
      $\kappa \upsilon$& Average number of effective UEs inside a LEO BS $-3$ dB beam cell.& $3$ \\
      $n$& Number of served UEs in each beam.&  $n=\kappa \upsilon=3$ \\
      $\tilde{\kappa} $& $\kappa/\log(2)$. \\
      $\tau$ & Min. SIR for a successful decoding.&$0.2=-7$ [dB] \\
      $\theta$ & SIR of the served UE.& \\
     \bottomrule
    \end{tabular}
  \end{center}
\end{table}


\begin{figure}[ht]
  {\includegraphics[width=\linewidth]{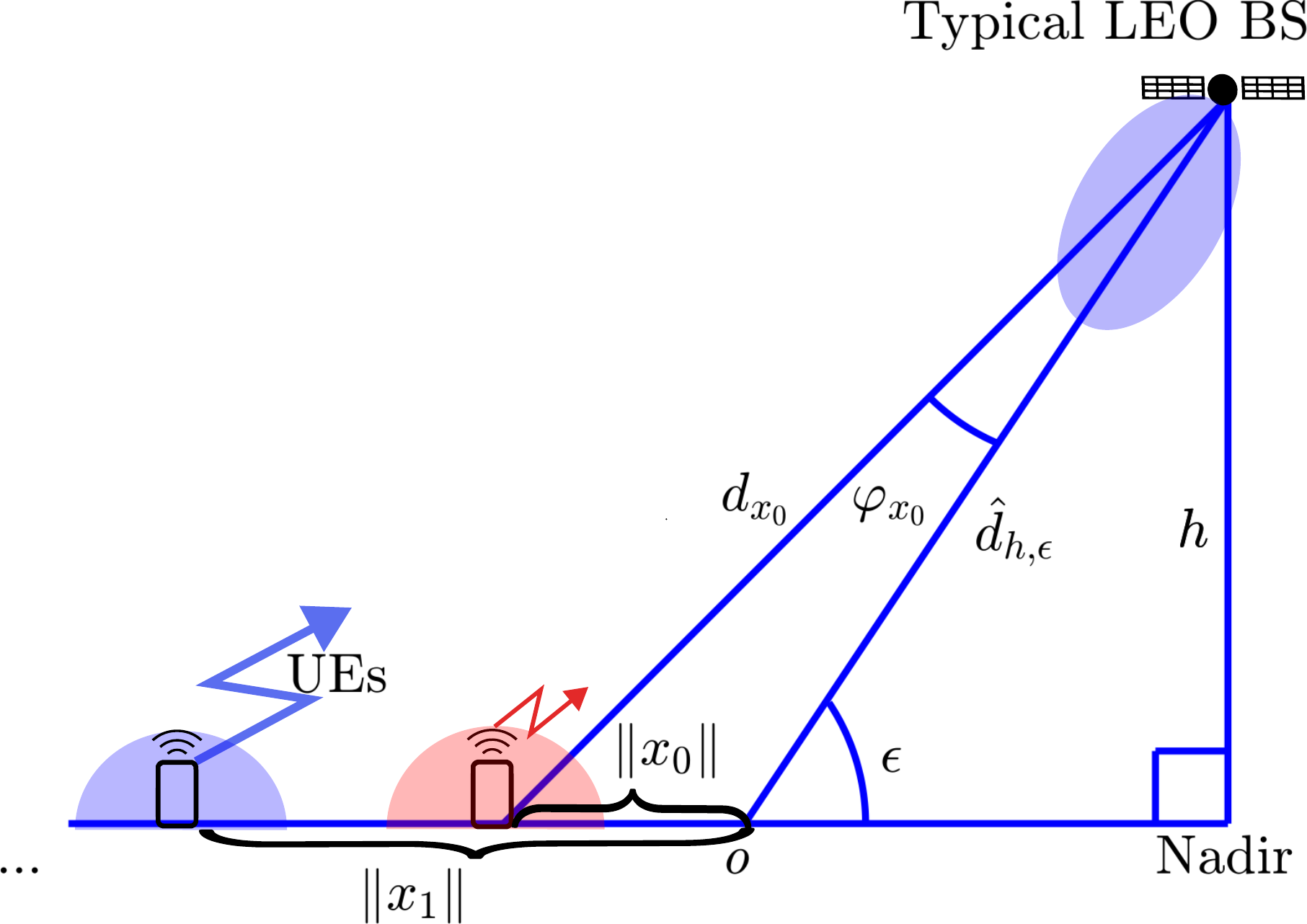}}
  \caption{A sketch of the planar system model. In the figure, the UEs $\{x_i\}$ are projected into the line according to their distance from the origin. The transmitter with the strongest signal is the first-served UE at the typical LEO BS beam}
    \label{fig:systemmodel}
\end{figure}

\label{sec:analysissec}

A narrowband LEO uplink is considered with narrow-beamed LEO receivers (beam half-width less than $4^{\circ}$ \cite{10909705}), and statistics over a short use period are studied. The LEO BSs form a homogeneous point process (p.p.) on a plane representing the constellation. Each LEO BS is at altitude $h$. The UEs follow a homogeneous PPP $\Phi \subset \R^2$ of density $\lambda$. The fading model is log-normal shadowing. Each LEO BS beam is associated with $n>0$ served UEs ordered according to their signal powers after the shadowing. $n$ is perceived as the average number of \textit{effective} UEs inside the beam cells, which is defined as the $-3$ dB footprint of the beams. The term ``effective'' means the fraction of transmitters that are ``not entirely'' attenuated due to shadowing; we will define this notion precisely in the work. For low elevation angles, when the footprint size grows and shadowing becomes increasingly prominent, the number of effective UEs is kept constant by decreasing the beam half-width $\varphi_{\text{RX}}$ by the appropriate amount, varying from $\varphi_{\text{RX}} \in (0.6^{\circ},1.6^{\circ})$ for the respective beam center elevation angles $\epsilon \in (30^{\circ},90^{\circ})$ for $n=3$. The crucial parameter is theoretically $n$, while the other parameters only slightly affect the simulated values in the spherical model. Please refer to Table \ref{table:parameters} for the numerical parameters used in the results section simulations. 

Figure \ref{fig:systemmodel} depicts the planar system model represented on the line. Because of the translation invariance of the PPP, all locations are statistically equivalent, and we define the origin $\textit{o}$ to represent the typical LEO BS beam center. Furthermore, we will demonstrate how the statistics only depend on the number of effective UEs inside the beam cells; hence, we can refer to a ``typical beam'' if the effective number of UEs is fixed. We assume a fast-decaying narrow beam; hence, each distance $\hat{d}_{h,\epsilon} \approx {d}_{x}$ can be approximated as equal for all (relevant) UEs. Hence, the spatial path losses cancel in the SIR in the analytical model. Furthermore, without the side lobes, the same notion applies to the transmission powers.

The antenna gain at the typical LEO BS beam is given for the Euclidean distance $\|x\| \in [0, \infty)$ as a Gaussian function
  \begin{equation}
    \label{eq:Gaussianantpat}
    G(\|x\|) = 2^{-(D_{h,\epsilon}\|x\|)^2 / \varphi_{\text{RX}}^2},
  \end{equation}
  where the angle $\varphi_{\text{RX}}$ denotes the $-3$ dB antenna gain width. The scaling constant $D_{h,\epsilon} \triangleq \sin^2(\epsilon)/h$ is a first-order coefficient of the Taylor expansion of the angle $\varphi_x$ w.r.t. the boresight of the typical antenna pattern (see extensive details in \cite{10909705}).



  \subsection{Spherical system model}
  All analytical results based on the planar model are tested against Monte Carlo simulations based on a system model with a spherical Earth. In the spherical model, all angles and distances are calculated according to spherical geometry. For a more detailed analysis and comparison of the planar model with the spherical model, please refer to \cite[Fig. 4]{10909705}. For narrow beams, the proposed analytical planar model is highly accurate for beam widths smaller than $\varphi_{\text{RX}}<4^{\circ}$. The number of Monte Carlo simulations corresponds to stable statistical convergence.
  
  \subsection{Shadowing}


\subsubsection{Gaussian mixture shadowing model (simulations)}
\label{sec:guassianmixture}

Consider a two-tier $\{\text{LoS},\text{NLoS}\}$ (line-of-sight and non-line-of-sight) Gaussian mixture shadow fading model with the parameters $\mu_{\text{LoS}} = 0$ dB, $\sigma_{\text{LoS}} = 4$ dB, $\mu_{\text{NLoS}} = -26$ dB, and $\sigma_{\text{NLoS}} = 6$ dB, which correspond to an urban environment \cite{TR38.811}. Assuming i.i.d. power shadow fading for all UEs, \textit{the typical shadowed transmit power} $H_{\mathcal{M}\mathcal{L}\mathcal{N}}$ follows a log-normal mixture distribution;
\begin{align}
  \label{eq:tier2lognormal}
  H_{\mathcal{M}\mathcal{L}\mathcal{N}} &\sim p_{\text{LoS}} \mathcal{L}\mathcal{N}(\rho \mu_{\text{LoS}}, (\rho \sigma_{\text{LoS}})^2) \nonumber \\
  &\quad + p_{\text{NLoS}} \mathcal{L}\mathcal{N}(\rho \mu_{\text{NLoS}}, (\rho \sigma_{\text{NLoS}})^2),
\end{align}
where $p_{\text{LoS}}=1-p_{\text{NLoS}}$ is the LoS probability. Considering a natural base for the log-normal distribution, the constant $\rho \triangleq \log(10)/10$ normalizes the parameters $\mu_{\text{LoS}}, \sigma_{\text{LoS}}, \mu_{\text{NLoS}},$ and $\sigma_{\text{NLoS}}$, ensuring that the conditioned r.v.'s $10 \log_{10}(H_{\mathcal{MLN}}|\text{LoS})$ and $10 \log_{10}(H_{\mathcal{MLN}}|\text{NLoS})$ evaluate to r.v.'s following the normal distributions $\mathcal{N}(\mu_{\text{LoS}}, \sigma_{\text{LoS}}^2)$ and $\mathcal{N}(\mu_{\text{NLoS}}, \sigma_{\text{NLoS}}^2)$, respectively.

\subsubsection{Mixture exponential shadowing distribution (analysis)}
As a compromise between analytical tractability and realism, we introduce an exponential distribution with an atomic probability measure at $0$, described by the complementary cumulative distribution function (CCDF)
\begin{equation}
  \label{eq:defexp}
  F_{{H}_{\text{Exp}}}(t) = \upsilon e^{-t} \text{ for }t\geq 0.
\end{equation}
 The parameter $0 \leq 1-\upsilon < 1$ denotes the probability that the shadowed signal is entirely attenuated and takes the value of zero, otherwise, the power follows the exponential distribution. 

\begin{figure}[ht]%
  \includegraphics[width=\linewidth]{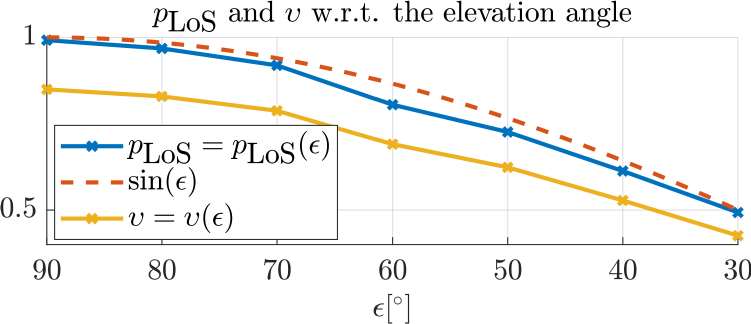}     
            \caption{The LoS probabilities and $\upsilon$ in the urban scenario}  \label{fig:upsilonvsepsilon} 
\end{figure}

In the following, the scaling term $\Upsilon$ ensures that the means of the log-normal mixture distribution and the mixture exponential distribution match: $\mathbb{E}(\Upsilon \times H_{\mathcal{MLN}}) = \mathbb{E}(H_{\text{Exp}}) = \upsilon$. By equating the first two moments M1 and M2 of $H_{\text{Exp}}$ (l.h.s.) and $\Upsilon_{}\times H_{\mathcal{M} \mathcal{L}\mathcal{N}}$ (r.h.s.);
\begin{equation}
  \label{eq:matchingmoments}
  \begin{cases}
    &\textup{M1: }\upsilon_{} = \Upsilon_{} \left(p_{\text{LoS}} e^{\mu_{\text{LoS}} + \sigma_{\text{LoS}}^2/2} + p_{\text{NLoS}} e^{\mu_{\text{NLoS}} + \sigma_{\text{NLoS}}^2/2}\right)\\
    &\textup{M2: }2\upsilon_{}= \Upsilon_{}^2 \left( p_{\text{LoS}} e^{2(\mu_{\text{LoS}} + \sigma_{\text{LoS}}^2)} + p_{\text{NLoS}} e^{2(\mu_{\text{NLoS}} + \sigma_{\text{NLoS}}^2)} \right), 
  \end{cases}
\end{equation}
 we can solve for the parameter $\upsilon$ for the shadowing fading distribution approximation $F_{H_{\textup{exp}}}(\cdot)$:
\begin{align}
  \label{eq:upsilon}
  & \upsilon_{}=\frac{ 2\left( p_{\text{LoS}}e^{\mu_{\text{LoS}}+\sigma^2_{\text{LoS}}/2}+p_{\text{NLoS}}e^{\mu_{\text{NLoS}}+\sigma^2_{\text{NLoS}}/2} \right)^2}{p_{\text{LoS}}e^{2(\mu_{\text{LoS}}+\sigma_{\text{LoS}}^2)}+p_{\text{NLoS}}e^{2(\mu_{\text{NLoS}}+\sigma_{\text{NLoS}}^2)}}.
\end{align}
The parameter $\upsilon=\upsilon(\epsilon)$ varies with the elevation angle through the LoS probability, which influences the shadow fading characteristics. We refer to $\upsilon$ as the \textit{effective UE} fraction. The parameter $\Upsilon_{}$ holds no significance in an interference-limited scenario, as the equal scaling of all UE powers neutralizes its effect. The urban scenario LoS probabilities are found in \cite[Sec. 6.6.1]{TR38.811}. The LoS probabilities and $\upsilon$ are plotted in Figure \ref{fig:upsilonvsepsilon} w.r.t. the elevation angle. 

\section{Analysis}

Given i.i.d. shadowing variables $\{H_x\}_{x \in \Phi}$, we define the process of the received signal powers at the typical LEO BS, referred to as the gain process (GP), by
\begin{equation}
  \label{eq:gainprocess}
  \mathcal{G} \triangleq \left\{ H_x G(\|x\|) : x \in \Phi \right\},
\end{equation}
where $\|x\|$ is the Euclidean distance from $\textit{o}$. 

The GP is a \textit{projection process} mapping the points from $\mathbb{R}^2$ into $(0,\infty)$ and, as such, forms a nonhomogeneous PPP \cite[Section 4.2.5]{alma998193414406526}.

Since the variables $\{H_x\}_{x \in \Phi}$ are i.i.d., we can denote the typical shadowing variable simply as $H$ without the subscript.
\begin{prop}[Density of the GP]\
  Let $F_H(\cdot)$ be the (possibly degenerate) CCDF of a fading variable $H$. The density function of $\mathcal{G}$ is given by
  \begin{equation}
    \label{eq:GPdensity}
    \lambda_{\mathcal{G}}(t) = \tilde{\kappa} {F_H(t)}/{t}, \quad t \in (0, \infty),
  \end{equation}
  where $\tilde{\kappa} = {\kappa}/{\log(2)}$ and
  \begin{equation}\kappa \triangleq \pi \lambda \left(\frac{\varphi_{\textup{RX}}h}{\sin^2(\epsilon)}\right)^2
    \label{eq:kappa}
  \end{equation}
  is (approximately) the average number of UEs inside a $-3$ dB beam. Here, the angles are in radians.

  \begin{proof}
    Let $f_H(\cdot)$ be the probability density function (PDF) of $H$. Denote $G^{-1}(\cdot)$ as the generalized inverse of $G$, defined as $G^{-1}(y) = \inf \{x : G(x) < y\}$. According to \cite[Eq. (4.55)]{alma998193414406526},
    \begin{align*}
      &\int_t^{\infty} \lambda_{\mathcal{G}}(y) \, dy = \pi \lambda \mathbb{E}\left[ \left({G^{-1}(t/H)}{}\right)^2 \right] \\
      &= \pi \lambda \int_t^{\infty} \left(-\frac{\varphi_{\text{RX}} \sqrt{-\log(t/h)}}{D_{h,\epsilon} \sqrt{\log(2)}}\right)^2 f_H(h) \, dh \\
      &= -\tilde{\kappa} \int_t^{\infty} \log(t/h) f_H(h) \, dh \\
      &\overset{(a)}{=} -\tilde{\kappa} \left[ \left. \log(t/h) F_H(h) \right|_t^{\infty} + \int_t^{\infty} \frac{F_H(h)}{h} \, dh \right].
    \end{align*}
    In (a), we use integration by parts. The result follows by differentiating with respect to $t$ and applying the negative sign. Note that a necessary condition for this procedure is that $\int_t^{\infty} \log(t/h) f_H(h) \, dh$ converges for all $t > 0$.
    
    Please refer to \cite[Lemma 1]{10909705} for throughout explanation of the geometric interpretation of $\kappa$.
  \end{proof}
\end{prop}
The total interference, or total received power, is defined as the sum of the GP at the typical LEO BS:
\begin{equation}
  \label{eq:totpow}
  I \triangleq \sum_{x \in \Phi} H_x G(\|x\|) = \sum_{x \in \mathcal{G}} x.
\end{equation}

The mean and the variance of $I$ are respectively given by
\begin{align}
  \label{eq:totmean}
  &\mathbb{E}\left(I \right) = \int_{0}^{\infty} t\lambda_{\mathcal{G}}(t) dt = \tilde{\kappa} \int_{0}^{\infty}F_H(t) dt =\tilde{\kappa} \mathbb{E}(H),\\
  \label{eq:totvar}
  &\text{Var}\left(I \right) = \int_{0}^{\infty} t^2\lambda_{\mathcal{G}}(t) dt= \tilde{\kappa} \int_0^{\infty}tF_H(t) dt  \nonumber \\
  &= \tilde{\kappa} \frac{\text{Var}(H) + \mathbb{E}(H)^2}{2} = \tilde{\kappa}  \mathbb{E}[H^2]/2.
\end{align}

Note that matching the first two moments of the fading distributions \eqref{eq:tier2lognormal} and \eqref{eq:defexp} is equivalent to matching the mean and the variance of the total interference.

\subsection{Laplace transform of the total interference}

Setting $v(x) = e^{-sx}$ in \cite[Eq. (14)]{10909705} yields
\begin{align}
  \label{eq:lapdef}
  &\mathcal{L}_{I}(s)\triangleq \mathbb{E}\left(e^{-sI}\right)\overset{(a)}{=} \exp\left\{-\int_0^{\infty}(1-e^{-sr}) \lambda_{\mathcal{G}}(r) dr \right\} \nonumber \\
  &=\exp\left\{-\tilde{\kappa}\int_0^{\infty}(1-e^{-sr}) F_{{H}_{\text{exp}}}(r) /r dr \right\} \nonumber \\
  &=\exp\left\{-\tilde{\kappa}\upsilon_{}\int_0^{\infty}(1-e^{-sr}) e^{-r} /r dr \right\} =(1+s)^{-\tilde{\kappa}\upsilon_{}},
\end{align}
which is the Laplace transform of the gamma distribution with the shape parameter $\tilde{\kappa}\upsilon_{}$.

    \subsection{Order statistics of the STIR and SIR processes}

 At the typical LEO BS, we denote the SIR process of the UEs as follows:
\begin{align}
  \label{eq:SINR}
  \Psi &= \{\mathsf{Z}: \mathsf{Z} \in \Psi\} \triangleq \left\{ \frac{u}{I-u} : u \in \mathcal{G}\right\} \nonumber \\
  &=\left\{ \frac{H_x G(D_{h,\epsilon}\|x\|)}{I-H_x G(D_{h,\epsilon}\|x\|)} : x \in \Phi\right\},
\end{align}
where $I$ is defined in \eqref{eq:totpow}. Similarly, the signal-to-total-interference ratio (STIR) process is defined as
\begin{align}
  \label{eq:STINR}
  \Psi' &= \{\mathsf{Z}': \mathsf{Z}' \in \Psi'\} \triangleq \left\{ \frac{u}{I} : u \in \mathcal{G}\right\}.
\end{align}
We can always recover the process from another:
\begin{equation}
  \label{eq:STINRSIRrealations}
  \Psi = \left\{ \frac{\mathsf{Z}'}{1- \mathsf{Z}'}: \mathsf{Z}' \in \Psi' \right\}, \hspace{0.3cm} \Psi' = \left\{ \frac{\mathsf{Z}}{1+ \mathsf{Z}}: \mathsf{Z} \in \Psi \right\}.
\end{equation}
Let $\theta$ denote the SIR threshold for successful transmission. The event $\Psi \ni\mathsf{Z}> \theta$ is equivalent to $\Psi' \ni \mathsf{Z}'> \theta'$  with $\theta' \triangleq \theta/(1+\theta)$ and $\theta \triangleq \theta'/(1-\theta')$. 
    
We denote $\mathsf{Z}'_{(1)}>\mathsf{Z}'_{(2)} >\mathsf{Z}'_{(3)} \dots$ as the order statistics of the STIR process $\Psi'$, such that $\mathsf{Z}'_{(1)}$ is the largest value in $\Psi'$. Through the monotonicity of the relations \eqref{eq:STINRSIRrealations}, the order statistics of the STIR process are equivalent to the order statistics of the SIR process.

\begin{prop}
  The density of the $k$th factorial moment measure \cite[Corollary 10]{7305791} of the STIR process at the typical LEO BS with a narrow Gaussian antenna beam under the mixture exponential shadowing is given by
  \begin{align}
    \label{eq:factorialmoment}
    \mu'^{(k)}(t_1',\dots,t'_k) = (\tilde{\kappa}\upsilon_{})^k\prod_{j=1}^k{t'}_{j}^{-1}\left(1- \sum_{j=1}^kt'_j \right)^{\tilde{\kappa}\upsilon_{}-1},       
  \end{align}
  whenever $t'_1>\dots >t'_k$ and $\sum_{i=1}^k t'_i \leq 1$, and $0$ otherwise.
  \begin{proof}
    By \eqref{eq:lapdef}, the total interference can be characterized by the gamma process at time $\tilde{\kappa}\upsilon$ \cite[Eq. (8)]{pitman1997two}. Hence, the STIR process $\Psi'$ can be characterized by a Poisson-Dirichlet distribution PD$(0,\tilde{\kappa}\upsilon)$ that has the given density \cite[Eq. (2.3)]{handa2009two}.
  \end{proof}
\end{prop}

The partial densities can be derived from the density of the $k$th factorial moment measure as \cite[Eq. (62)]{7305791}
\begin{align}
  \label{eq:auxillary}
  &{\mu'}_k^{(k+i)}(z'_1,\dots,z'_k) \nonumber \\
  &= \int_{z'_k}^1 \dots \int_{z'_k}^1 {\mu'}^{(k+i)}(z'_1,\dots,z'_k,\zeta'_1,\dots,\zeta'_i) d\zeta'_1 \dots d\zeta'_i,
\end{align}
the support of the density being in the region $\sum_{j=1}^nz'_j+iz'_k \leq 1$.

The joint PDF of the $k$ strongest values of the STIR process $(\mathsf{Z}'_{(1)}, \dots, \mathsf{Z}'_{(k)})$ is given as a series expansion involving the partial densities \cite[Eq. (64)]{7305791}:
\begin{equation}
  \label{eq:jointprobability}
  f'_{(k)}(z'_1,\dots,z'_k)= \sum^{i_{\text{max}}}_{i=0}\frac{(-1)^i}{i!}{\mu'}_k^{(k+i)}(z'_1,\dots,z'_k),
\end{equation}
for $z'_1>z'_2>\dots>z'_k$ and $f'_{(k)}(z'_1,\dots,z'_k) =0 $ otherwise. The upper bound for the index $i_{\text{max}}<1/z'_k-k$ corresponds to the non-zero terms of the series expansion. 

Directly from the joint PDF, the $k$-coverage probability that the first $k$ strongest signals reach the threshold $\theta$ is given by
\begin{align}
  \label{eq:kprobability}
  &\mathcal{P}^{(k)}(\theta) \triangleq  \int_{\theta'}^1\dots \int_{\theta'}^1 f'_{(k)}({z'_1},\dots,{z'_n})dz'_1 \dots d{z'_k}, 
\end{align}
with $\theta'=\theta/(1+\theta)$ and $i_{\text{max}}<1/\theta'-k$.

    The density of the $k$th factorial moment measure of the SIR process can be extracted from $\mu'^{(k)}$ \cite[Corollary 6.1.3]{alma998193414406526}:
    \begin{align}
      \label{eq:densitySINR}
      &\mu^{(k)}(z_1,\dots,z_k)&  \nonumber\\
     &= \prod_{j=1}^k\frac{1}{(1+z_j)^2}\mu'^{(k)}\left(\frac{z_1}{1+z_1},\dots,\frac{z_k}{1+z_k}\right).
    \end{align}

\subsection{SIR under interference cancellation}

Let $(u_{(1)}, \dots, u_{(k)}) \subset \mathcal{G}$ represent an ordered set of points in the GP, where $u_{(1)}$ denotes the strongest signal at the typical LEO BS beam. Conditioned on the first $k-1$ signals being successfully removed from the total interference, under the SIC, we denote the SIR of the $k$th strongest UE by
\begin{equation}
  \label{eq:IC-SINR}
  \text{SIR}_{(k)} \triangleq \frac{u_{(k)}}{I-\sum_{j =1}^k u_{(j)}}.
\end{equation}

Let us first study $\text{SIR}_{(1)}$. Combining \eqref{eq:factorialmoment}, \eqref{eq:jointprobability}, and \eqref{eq:densitySINR}, we can derive a closed-form for the SIR PDF of the strongest signal in the \textit{simple coverage region} $z\geq 1$: $f_{(1)}(z) =  {\tilde{\kappa}\upsilon_{}\left({z + 1} \right)^{-\tilde{\kappa}\upsilon_{}}}/{z}$~\footnote{The SIR has a heavy-tailed distribution, cf. \cite[Eq. (32)]{10909705}.}. The first and the second moments of the SIR are bounded by
\begin{align}
  \label{eq:SIR1}
  &\mathbb{E}(\text{SIR}_{(1)}) \geq \int_{1}^{\infty}f_{(1)}(z) z dz=\frac{  2^{1-\tilde{\kappa} \upsilon } \tilde{\kappa} \upsilon }{\tilde{\kappa} \upsilon -1}, \\
&\mathbb{E}(\text{SIR}^2_{(1)}) \geq \int_{1}^{\infty}f_{(1)}(z) z^2 dz = \frac{2^{1-\tilde{\kappa}v}(\tilde{\kappa}v)^2}{(\tilde{\kappa}\upsilon-1)(\tilde{\kappa}\upsilon-2)},  \label{eq:SIR2}
\end{align}
respectively. These are divergent for $\tilde{\kappa}\upsilon \leq 1$ and $\tilde{\kappa}\upsilon \leq 2$, respectively: for less than $2\log(2)$ effective UEs inside the beams on average, the first and second moments---hence, also the variance---are infinite (or undefined). Despite the strong average SIR, the infinite variance for $\tilde{\kappa}\upsilon \leq 2$ is not desirable if we want ``good'' user fairness, explicitly defined as finite SIR variance in this paper.

We have the following identity in terms of the STIR process \cite[Eq. (69)]{7305791}:
\begin{equation}
  \label{eq:identityintermsoftheSTIR}
   \mathbb{P}(\text{SIR}_{(k)} > \theta) = \mathbb{P}\left(\mathsf{Z}'_{(k)}+\theta'\sum_{j=1}^{k-1}\mathsf{Z}'_{(j)} > \theta'\right).
\end{equation}



Finally, we consider the SIR under the \textit{perfect} successive signal cancellation (SIC-SIR).~\footnote{Imperfect interference cancellation is straightforward to analyze by adding an additional term in $(0,1)$ \cite[Eq. (6.21)]{alma998193414406526} to \eqref{eq:identityintermsoftheSTIR}.} Under the SIC, a necessary condition for the successful reception of the $k$th strongest UE at the typical LEO BS is that the preceding $k-1$ signals are successively decoded and removed from the interference. We denote the minimum SIR threshold for a successful decoding by $\tau$ and the corresponding $\tau' = \tau/(1+\tau)$. In this work, we set $\tau = -7$ dB. Formally, the conditions for the SIC-SIR of the $k$th UE to exceed $\theta \geq \tau$ are \begin{enumerate}
\item $ \mathsf{Z}_{(m)}'+\tau'\sum_{j=1}^{m-1}\mathsf{Z}_j'>\tau'$ { for all } $m \in \{1, \dots, k-1\}$,\textup{ and} 
  \item $\mathsf{Z}_{(k)}'+\theta'\sum_{j=1}^{k-1}\mathsf{Z}_j'>\theta'.$
\end{enumerate}  We can use the order statistics theory to study the coverage probability under the SIC. 

\begin{prop}[SIC-SIR]
  Under the SIC, the coverage probability of the UE with $k$th strongest signal is given by
  
  \begin{align}
    \label{eq:SICprob}
     &\mathcal{P}^{(k)}_{\textup{SIC}}(\theta)\triangleq \int_{0}^{1} \dots \int_{0}^{1}d z'_1 \dots  d z'_k \times f'_{(k)}(z'_1,\dots,z'_k)       \\
    &  \times \mathds{1}\left( z'_k+ \theta'\sum_{j=1}^{k-1} z'_j>\theta'  \right)\prod_{m=1}^{k-1} \mathds{1}\left( z'_m+ \tau'\sum_{j=1}^{m-1} z'_j>\tau'  \right)\nonumber
  \end{align}
  with the upper summation limit in $f'_{(k)}(\cdot)$ bounded by $i_{\text{max}} < 1/\tau'-1=1/\tau$. $\mathds{1}(\cdot)$ is the indicator function. 
  \begin{proof}
    The expression follows using the joint PDF of the order statistics \eqref{eq:jointprobability}.  $i_{\text{max}}$ can be relaxed by the condition $z'_{k}+\tau'\sum_{j=1}^{k-1}z'_{j}>\tau'$. By simple algebra, $\sum_{j=1}^{k-1}z_{j}> 1-z_{k}/\tau'$. Recall the condition on the non-zero terms of $\mu'^{(k+i)}$:  $\sum_{j=1}^k z'_{j}+i z'_{k} =\sum_{j=1}^{k-1}z'_{j} +z'_{k}+i z'_{k}  \leq 1$. The condition certainly does \textit{not} hold if $1-z_{k}/\tau'+ z'_{k}+i z'_{k}>1$. We arrive at the inequality $z'_{k} \left(-1/\tau' + 1 +i \right)>0$. Divide both sides by $z'_{k}>0$, and the general upper bound of $i$ follows.
  \end{proof}
\end{prop}

The expected total bandwidth-normalized throughput \cite[Eq. (7.19)]{alma998193414406526} of the $n$ served UEs under the SIC (see analogous derivation in \cite[Eq. (36)]{10909705})
is given by
\begin{align}
  \mathcal{T}_{n}&\triangleq \sum_{k=1}^n \mathbb{E}\left(\frac{\log\Big(1+\text{SIR}_{(k)}\prod_{i=1}^{k-1}\mathds{1}(\text{SIR}_{(i)}>\tau)\Big)}{\log(2)}\right) \nonumber \\
  &=\sum_{k=1}^n\int_{\tau}^{\infty}\frac{\mathcal{P}^{(k)}_{\textup{SIC}}(t)}{(t+1)\log(2)}d t \textup{ (bit/s/Hz)}. \label{eq:throughput}
\end{align}

\section{Numerical results}
  
\begin{figure}[ht]
  \centering
  \includegraphics[width=\linewidth]{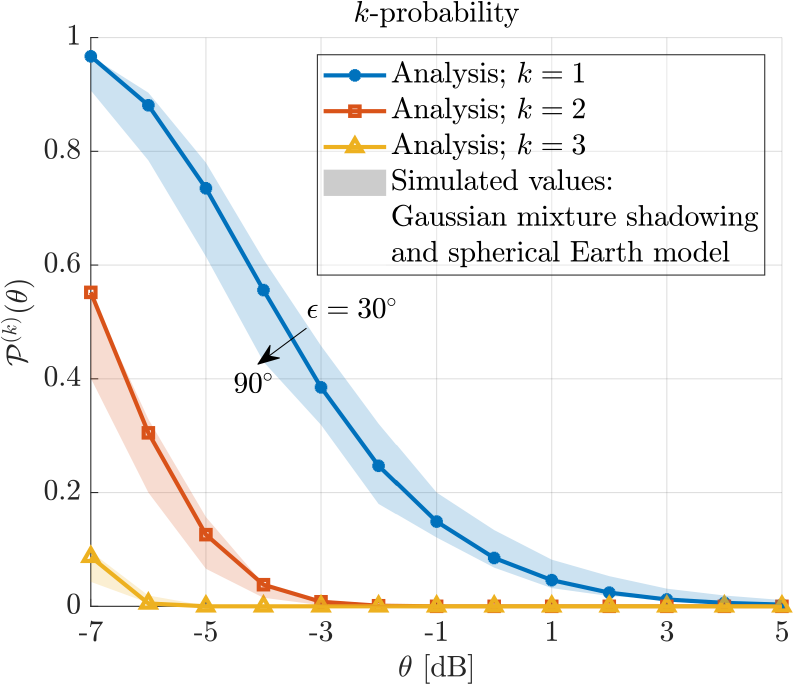}
  \caption{The SIR $k$-probabilities of the UEs $x_{(k)} \in \{x_{(1)},x_{(2)},x_{(3)}\}$ with $\kappa v=3 $ (on average, three effective UEs inside each beam cell)
} 
  \label{fig:nprobability}
\end{figure}

\begin{figure}[ht]
  \centering
  \includegraphics[width=\linewidth]{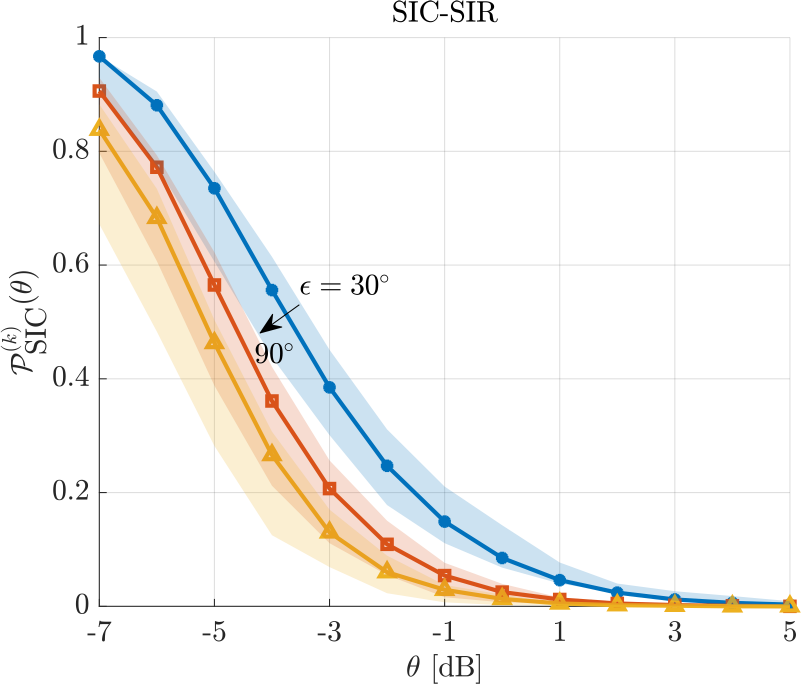}
  \caption{The SIC-SIR of the UEs $x_{(k)} \in \{x_{(1)},x_{(2)},x_{(3)}\}$ with $\kappa v=3 $ (on average, three effective UEs inside each beam)} 
  \label{fig:ASICSIR}
\end{figure}

Figures \ref{fig:nprobability} and \ref{fig:ASICSIR} depict the $k$-probabilities \eqref{eq:kprobability} and the SIC-SIR \eqref{eq:SICprob} for ${\kappa}\upsilon=\tilde{\kappa}\upsilon\log(2)=n=3$, respectively. The system parameters are presented in Table \ref{table:parameters}. In the simulations, $\varphi_{\text{RX}}$ is determined to match ${\kappa}\upsilon={\kappa}(\varphi_{\text{RX}},\epsilon)\upsilon(\epsilon)=3$ for each $\epsilon \in (30,90)\degree$ so that the number of effective UEs inside each LEO BS beam is equal. Explicitly, the beam width can be derived according to \eqref{eq:kappa} and is given by $ \varphi_{\text{RX}}(\epsilon) = \sqrt{\upsilon(90 \degree)/\upsilon(\epsilon)} \times 1.6 \degree \times\sin^2(\epsilon) \in (0.6,1.6) \degree $. 

The performance of the $2$nd and $3$rd UEs is significantly enhanced with the SIC-SIR compared to the $k$-probability, especially for small $\theta$ (i.e., low-rate/throughput transmissions), showing a $((0.8620-0.0870)/0.0870) \times 100 \% \approx 900 \%$ improvement for the coverage probability of the third UE with $\theta = -7$ dB.

$ n= \kappa \upsilon = 3$ is in the finite-variance region of each transmitter's SIR distribution (recall \eqref{eq:SIR2}), i.e., in the good user fairness region. Hence, the link quality is more stable at the LEO BS than in a lower-UE-density scenario, say $n=\kappa \upsilon =1$, where the SIR variance is infinite. A loose upper bound for the variance of the strongest UE SIR in Figures \ref{fig:nprobability} and \ref{fig:ASICSIR} can be calculated in accordance with \eqref{eq:SIR1} and \eqref{eq:SIR2}: $\text{var}(\text{SIR}_{(1)}) =\mathbb{E}(\text{SIR}^2_{(1)})-\mathbb{E}(\text{SIR}_{(1)})^2  < 1.75$. In this regard, SIC-SIR enables more consistent UE-LEO BS link quality by allowing consistent UE performance. Next, we study the total throughput.

Considering the SIR distributions in Figure \ref{fig:ASICSIR}, using the trapezoidal method, we can calculate the throughput \eqref{eq:throughput} as $\mathcal{T}_{3}\approx 0.7 \textup{ bit/s/Hz}$. On the other hand, in the infinite-variance SIR region with one served UE in each cell on average (i.e., $n=1$), the served transmitter achieves $\mathcal{T}_1 \approx 0.9$ bit/s/Hz. That is, under the SIC with $n=3$, there is a considerable but not overwhelming $-0.2$ bit/s/Hz tradeoff in total throughput and user fairness. Smaller $\tau$ yields a better tradeoff. $n=2$ yields similar throughput to $n=1$ and is in the good user fairness region, with an upper bound for the SIR variance $\text{var}(\text{SIR}_{(1)}) < 2$.


\section{Conclusion}
We developed a stochastic geometry analysis of SIR order statistics and analyzed NOMA LEO uplink performance under SIC. The analytical results were validated against realistic Monte Carlo simulations. The analytical framework is highly tractable, leveraging closed-form expressions for the interference and SIR distributions, which provide insight into link performance that goes beyond numerical simulations.

SIC was observed to improve user fairness while maintaining good average total throughput across LEO BS beams, providing design recommendations for next-generation LEO networks. On the downside, with multiple users in a beam, the signal quality of each UE-LEO BS link is rather poor, even though the aggregate throughput is good. Furthermore, the second- and third-strongest UEs receive worse link quality than the strongest UE. Further studies are required to clarify the associated trade-offs and implementation considerations.


\bibliographystyle{IEEEtran}
\bibliography{IEEEabrv,source}

\end{document}